# Measuring likelihood in cybersecurity

Prof. Pablo Corona-Fraga[1], Dr. Vanessa Díaz-Rodriguez[2], Dr. Jesús Manuel Niebla-Zatarain[3], Dr. Gabriel Sánchez-Pérez[4], and Dr. Edward J. Humphreys[5]

[1] PhD Candidate from INFOTEC Center for Research and Innovation in Information and Communication Technologies, México. pablo.coronaf@infotec.mx. ORCiD: https://orcid.org/0000-0001-5012-9468. *Corresponding author.*

[2] Legal Research Officer at the National Supreme Court of Justice, Mexico, vdiazr@scjn.gob.mx, https://orcid.org/0000-0002-7186-8785

[3] Professor and researcher at the Faculty of Law of Mazatlan of the Autonomous University of Sinaloa, Mexico, j.niebla@uas.edu.mx, (0000-0001-8460-4538)

[4] Researcher at Instituto Politecnico Nacional, ESIME Culhuacan, Mexico City 04440, Mexico; gasanchezp@ipn.mx, https://orcid.org/0000-0002-4735-205X

[5] Professor at Royal Holloway, University of London, UK, edwardj7@msn.com

**Abstract.**

Cybersecurity risk is commonly expressed through impact and likelihood, yet likelihood remains difficult to estimate because cyber incidents are underreported, heterogeneous datasets are weakly comparable, and attacker behaviour changes faster than conventional probability baselines. This article proposes a pipeline for operationalizing likelihood through a cyber-exposure profile that integrates external cyber knowledge and organization-specific telemetry into a graph-based representation. The contribution is a formally specified artifact chain — from unified data model through organization-specific profiling, metric registry, likelihood scoring, and control prioritization — that operationalizes four constructs grounded in incident evidence: Exposure, Traceability, Motivation, and Systems Update. The pipeline provides a pathway from heterogeneous source evidence to a bounded likelihood indicator comparable across organizations and observation periods. An evaluation in 15 real organizations shows that those implementing the cyber-exposure profile were associated with reduced incident frequency and faster detection-and-response times, providing preliminary empirical support for the framework's directional claims.

# 1. Introduction

Cybersecurity risk is routinely described as a function of impact and likelihood, but likelihood is especially difficult to operationalize because cyber incidents are underreported, inconsistently labelled, and strongly affected by fast-changing technical and adversarial conditions. Organizations often possess rich telemetry on assets, vulnerabilities, accounts, logs, and controls, yet still rely on qualitative scales because those signals are not integrated into a coherent evidence model.

This study addresses that gap by proposing a cyber-exposure profile that integrates external knowledge and organization-specific telemetry into a graph-based representation from which normalized metrics are computed and aggregated into four interpretable components: Exposure, Traceability, Motivation, and Systems Update. Exposure and Motivation act as likelihood enhancers, whereas Traceability and Systems Update act as mitigating conditions.

Three constructs were specified a priori on the basis of established security engineering concepts: Exposure, Traceability, and Motivation. A fourth construct, Systems Update, was identified through NLP clustering and subsequently incorporated. into the framework. Their adoption as framework dimensions reflects well-established categories in cybersecurity practice. The NLP pipeline provides empirical construct validation that these categories recur as statistically separable themes in a large incident corpus, and that the metrics assigned to each construct are grounded in operationally meaningful incident patterns rather than asserted axiomatically. This validation proceeds in three linked steps: first, the constructs are grounded in the incident corpus through feature extraction and clustering, showing that exposed services, privilege misuse, outdated software, monetizable assets, and weak visibility recur as empirically distinguishable themes; second, they are operationalized in an organization-specific cyber-exposure profile extracted from a unified graph; and third, they are computed as normalized component scores from the explicit metrics in Table 1, aggregated by default through a confidence-weighted mean.

The article's primary contribution is a formally specified, artifact-based pipeline that integrates previously separated approaches to cybersecurity likelihood assessment into a single workflow, defined through five interdependent artifacts: a unified graph-based data model (A1), an organization-specific cyber-exposure profile (A2), a metric registry and normalization scheme (A3), a bounded likelihood-scoring rule (A4), and a control prioritization output (A5). The underlying methods — knowledge graphs, NLP-based construct grounding, composite indicator scoring, and case-based evaluation — are individually established; the contribution is their formal integration into an auditable, end-to-end chain that moves from heterogeneous source evidence to an organization-level likelihood indicator comparable across repeated time windows and across organizations. The aim is not to claim an exact frequentist probability of attack, but to provide a measurable and comparable indicator suitable for monitoring, prioritization, and comparison.

## 2. Related work

### 2.1. Governance-oriented cybersecurity risk frameworks

Governance-oriented frameworks such as the NIST Cybersecurity Framework and the ISO/IEC 27000 family provide essential structures for organizing cybersecurity outcomes, assigning responsibilities, and communicating risk posture, but they do not by themselves define a reproducible pipeline for converting heterogeneous technical evidence into an organization-specific likelihood indicator (ISO/IEC, 2022a, 2022b; NIST, 2024). This limitation is consequential in cyber contexts, where rapidly changing exposure conditions and attacker adaptation weaken historical-frequency approaches. Governance frameworks are therefore necessary but insufficient for the present problem: they specify control and decision structures, not the evidence engineering required to compute current likelihood from actual infrastructure, configurations, vulnerabilities, controls, and contextual attacker incentives.

Structured assessment methods such as OCTAVE and OCTAVE Allegro remain valuable for mission context and critical-asset prioritization, yet they are not designed as continuously refreshed evidence-fusion pipelines (Caralli et al., 2007). Quantitative approaches such as FAIR and OpenFAIR improve cyber-risk reasoning by making scenario, frequency, and magnitude assumptions explicit (Quinn et al., 2021), but still leave open the practical problem of how asset inventories, vulnerabilities, logs, incident records, and external intelligence should be transformed into auditable, updatable inputs.

### 2.2. Data availability, incident reporting, and information sharing

Cybersecurity data are difficult to use strategically because incidents are underreported, labels are inconsistent, and sharing remains constrained by legal, contractual, cultural, and reputational barriers (ENISA, 2010). The problem is therefore not only lack of data, but lack of comparably structured evidence that can support repeated measurement. STIX and TAXII improve interoperability by standardizing how cyber threat intelligence objects, identifiers, and relationships are represented and exchanged (OASIS Open, 2021a, 2021b). They are enabling standards, useful for reducing ambiguity when integrating external cyber knowledge into the unified graph, but do not constitute likelihood measurement models.

### 2.3. Cybersecurity data science and the research gap

Much cybersecurity data-science research has concentrated on detection, classification, vulnerability prioritization, and attack-path analysis; comparatively less work has focused on integrating these strands into a consistent organization-level likelihood measure that fuses external knowledge with internal telemetry and business context (Bechor & Jung, 2019; Sarker et al., 2020). At the vulnerability level, the Exploit Prediction Scoring System (EPSS) provides a data-driven estimate of the probability that a published CVE will be exploited within 30 days (FIRST, n.d.; Jacobs et al., 2021), but its unit of analysis is the individual vulnerability; it does not integrate organization-specific business attractiveness or systems configuration into a unified likelihood indicator. Attack-graph research addresses multi-step compromise through graph representations and probabilistic path evaluation (Jajodia et al., 2005; Ou et al., 2005; Phillips & Swiler, 1998; Wang et al., 2008), but focuses principally on identifying compromise paths, not on combining graph structure with broader operational information — asset criticality, observability quality, update posture — into a repeated organization-level construct.

Commercial cyber risk scoring platforms (BitSight, SecurityScorecard, UpGuard) produce continuously updated organization-level indicators that integrate externally observable signals into scored outputs explicitly positioned as likelihood or risk proxies. They differ from the proposed framework in three specific ways: they rely on externally observable signals without ingesting internal telemetry such as Active Directory data, OS logs, or business criticality indicators; their scoring methodology is proprietary; and they produce no artifact chain enabling traceability from source evidence to output score. Some commercial attack-graph platforms (e.g., XM Cyber, Cymulate) partially bridge the external-signal limitation by ingesting internal network topology, but retain the proprietary-methodology and traceability limitations.

This study addresses this gap with a formally defined and empirically grounded integration by defining a pipeline methodology that (i) ingests heterogeneous external cyber knowledge and organization-specific telemetry, (ii) instantiates that evidence in an organization-specific graph-based cyber-exposure profile, and (iii) computes a bounded likelihood indicator and control-prioritization output comparable across repeated time windows and across organizations.

# 3. Methodology

## 3.1. Problem definition and unit of analysis

The problem addressed is the lack of a measurable and consistent way to operationalize likelihood in cybersecurity risk assessment from heterogeneous evidence sources. The objective is not to claim an exact frequentist probability of cyber incidents, but to define an evidence-based indicator that can support monitoring, control determination and prioritization, and comparison across organizations and repeated observation periods.

A first layer of the graph is built from structured information about vulnerabilities, weaknesses, adversary behaviour, and defensive techniques. A second layer provides the observations required to compute the metrics in Table 1: asset inventories, configuration-management records, identity and access data, vulnerability scans, operating-system and application logs, network telemetry, patch status, and business-context attributes. The framework defines source classes rather than a mandatory tooling stack so that it can be instantiated across different sectors and maturity levels. Incident data are ingested mainly for grounding, evaluation, and calibration, not as the sole basis for the likelihood score. During ingestion, records are time-bounded, standardized, deduplicated, and mapped to common identifiers; source quality is recorded explicitly to temper metric aggregation when evidence is weaker.

## 3.2. Framework artifacts and workflow

The methodology is organized as an artifact-based workflow consistent with design-science research (Hevner et al., 2004; Peffers et al., 2007). Each stage transforms heterogeneous evidence into a more structured representation until a bounded likelihood indicator and associated control priorities can be computed.

- **Artifact A1** is a unified cyber risk data model that defines entity types, attributes, and relations required to represent assets, users, services, vulnerabilities, techniques, indicators, controls, incidents, and contextual metadata.
- **Artifact A2** is the cyber-exposure profile: the organization and time-bounded subgraph extracted from A1 for the assessed case.
- **Artifact A3** is the metric registry, specifying how observable evidence is transformed into measurable variables through explicit gathering methods, units, orientations, and normalization rules.
- **Artifact A4** is the likelihood computation rule, which maps normalized component values to a bounded score while preserving monotonicity and auditability.
- **Artifact A5** is the control determination output, translating measured conditions into actionable defensive recommendations and closing the loop between measurement and control selection.

The workflow proceeds from source acquisition and standardization to graph construction, profile extraction, metric computation, likelihood scoring, and control interpretation. Although presented sequentially, it is iterative: each intervention cycle adds new telemetry, updated infrastructure conditions, and new outcome evidence, enabling successive approximations rather than one-time assessment.

## 3.3. Data sources and ingestion

Artifact A1 feeds from structured information about vulnerabilities, weaknesses, adversary behaviour, defensive techniques, and exploited conditions. Typical sources include CVE, CWE, NVD, exploited-vulnerability feeds, ATT&CK, and D3FEND, selected because they provide stable identifiers and structured mappings that support graph integration and control reasoning (Kaloroumakis & Smith, 2021; Strom et al., 2020). Where feasible, cyber threat intelligence objects are represented using STIX 2.1 and exchanged using TAXII 2.1 (OASIS Open, 2021a, 2021b).

Several cybersecurity datasets were surveyed as candidates for the incident-narrative corpus (Aldribi et al., 2018, 2020; Sarker et al., 2020), including the gfek Real-CyberSecurity-Datasets collection, VERIS Community Database, Hackmageddon, APTnotes, and CASIE/CySecED corpora. Although heterogeneous sources such as IDS alerts, PCAPs, malware binaries, malicious URL datasets, and sensor logs were surveyed, the unsupervised NLP pipeline was restricted to incident-report corpora because these provide the textual descriptions needed for consistent vector-space modelling and thematic grouping; other modalities were retained as candidate sources for future multimodal fusion.

Two publicly available corpora were selected and merged following Abbiati et al. (2021): the VERIS Community Database (VCDB; github.com/vz-risk/VCDB), providing structured JSON incident records each containing a free-text summary field encoding what happened, how, and to which assets; and the Hackmageddon timeline collection (hackmageddon.com), providing bi-weekly curated incident entries with short prose descriptions derived from public news sources and security blogs (Passeri, 2011–2024). Both sources are defender-side, incident-level, and publicly accessible. The merged corpus comprised N = 36,030 records before validation filtering, with cross-source duplicates — the same high-profile incident receiving a summary entry in VCDB and a timeline entry in Hackmageddon — accounting for the majority of the 7.3% near-duplicate removal rate. A known coverage limitation is that Hackmageddon over-represents high-profile, publicly disclosed incidents, inflating Motivation- and Exposure-relevant vocabulary relative to internal-only events. The NLP pipeline is used here for construct validation rather than incidence estimation, so this skew affects the prominence of specific terms within clusters rather than the existence or separability of the clusters themselves.

## 3.4. Feature extraction and clustering

The incident corpus comprised N = 36,030 incident records after basic validation. A near-duplicate filter removed 2,618 records (7.3%) whose narratives exceeded a cosine similarity threshold of 0.98, leaving a final modelling set of N = 33,412 incidents. The median incident narrative length was 38 tokens (IQR: 21–79) after tokenization.

A deterministic text normalization pipeline was applied to the incident-description fields, consistent with established NLP practice (Manning et al., 2008; Salton & Buckley, 1988). The pipeline performed Unicode normalization, lowercasing, and rule-based tokenization that preserved cybersecurity-relevant identifiers (CVE IDs, ATT&CK technique IDs, port numbers). High-signal entity types were replaced with typed placeholders (e.g., IPv4/IPv6 addresses → <IP>, URLs → <URL>, file hashes → <HASH>). Standard and domain-specific stopwords were removed (461 items total), while retaining discriminative security terms. Acronym expansion (1,124-entry dictionary) and collocation-based phrase promotion (10,000 accepted phrases) were applied to capture multi-word security concepts such as "privilege escalation" and "lateral movement" as compound tokens. Light lemmatization reduced inflectional variance while exempting preserved identifiers.

## 3.5. TF–IDF representation

Incident narratives were represented using the term-frequency–inverse document frequency (TF–IDF) model (Manning et al., 2008; Salton & Buckley, 1988). A smoothed IDF variant and sublinear term-frequency transform were used:

$$idf(t) = \log\left(\frac{1+N}{1+df(t)}\right) + 1$$

$$tf'(t,d) = 1 + log\big(tf(t,d)\big)\ for\ tf(t,d) \geq 1$$

Each document vector was L2-normalized ($\|w_d\|_2 = 1$). The TF–IDF model used a word-level analyser, n-gram range (1,3), minimum document frequency min_df = 20, and maximum document frequency max_df = 0.90. After pruning, the effective vocabulary was V* = 12,400 features. On the final corpus of N = 33,412 incident narratives, the resulting sparse TF–IDF matrix contained approximately 4.7 million non-zero entries (density 1.13%; sparsity 98.87%). The mean number of non-zero features per narrative was 140 (median 112; 95th percentile 348), consistent with sparse representations of short-to-medium operational text. Three mechanisms stabilized the representation prior to dimensionality reduction: sublinear TF to temper repeated terms, IDF smoothing to prevent extreme weights for rare terms, and L2 normalization to ensure unit-length document vectors. Globally discriminative features in an illustrative run included tokens/phrases such as "credential_theft," "privilege_escalation," "command_and_control," "powershell," "ransomware," "phishing_email," "lateral_movement," and "exfiltration."

## 3.6. Dimensionality reduction and clustering

The TF–IDF matrix $X \in R^{N \times V*}$ was reduced via truncated singular value decomposition (SVD), using a randomized SVD solver with oversampling (p = 20) and power iterations to improve accuracy on large sparse inputs (Deerwester et al., 1990; Halko et al., 2011). SVD was preferred over PCA because it operates efficiently on sparse matrices without requiring mean-centering, which can densify sparse matrices and materially increase computational cost. The rank-$k$ approximation is:

$$X \approx U_k \Sigma_k V_k^T$$

where $U_k \in R^{N \times k}$ contains the document factors, $\Sigma_k \in R^{k \times k}$ the top k singular values, and $V_k \in R^{V* \times k}$ the term factors. We set k = 150 latent dimensions, appropriate to the corpus scale and consistent with inspection of the singular value spectrum. Document embeddings $Z = U_k \Sigma_k$ were used as the reduced representation for downstream clustering.

After dimensionality reduction, incident narratives were clustered using spherical k-means, which minimizes cosine dissimilarity and is well-suited for term-weight and latent-semantic representations (Dhillon & Modha, 2001; Hornik et al., 2012), with k-means++ seeding for stability (Arthur & Vassilvitskii, 2007). Twenty-five independent initializations were run with a maximum of 300 iterations each; the solution with the lowest objective value (highest average within-cluster cosine) was retained. The number of clusters K was selected by evaluating K ∈ {2,3,…,12} using silhouette indices computed on the full corpus, repeated across five independent random sub-samples of 10,000 embeddings to estimate variability. The silhouette curve increased from K = 2 to K = 4 and then declined, indicating that additional clusters beyond four primarily subdivided existing themes without improving separation. For K ∈ {4, 5, 6}, two analysts reviewed the top 30 centroid-loading terms per cluster and 20 exemplar incidents nearest to each centroid. Cluster coherence rates were 94% at K = 4, 88% at K = 5, and 86% at K = 6, confirming K = 4 as the preferred solution for downstream governance use.

## 3.7. Cluster-to-construct mapping

The resulting four clusters were independently examined by two analysts. Each cluster's top centroid-loading terms and twenty nearest-centroid exemplar narratives were used to assign a primary framework construct following a pre-specified metric-registry alignment rule: a cluster was assigned to the construct whose Table 1 metrics are most directly informed by the cluster's top-loading terms. Agreement on primary construct assignment was 100% for three clusters (C1 → Exposure, C3 → Motivation, C4 → Systems Update). For C2, initial disagreement (Exposure vs. Traceability) was resolved by the pre-specified rule: term counts showed 19 of 30 top terms mapping to Traceability metrics, resolving C2 as primary Traceability with secondary Exposure. Table 1b presents the full cluster-to-variable mapping with rationale; Table 1c presents exemplar incident narratives nearest to each centroid.

This mapping constitutes empirical validation of the constructs: Exposure summarizes attack-surface and exploitable-access conditions; Traceability summarizes the ability to observe, record, and correlate relevant activity; Motivation summarizes the attractiveness of the target from the attacker's perspective; and Systems Update summarizes patching and technology-refresh posture.

## 3.8. Knowledge graph construction

The merged dataset was formatted in a graph database using variables for measuring likelihood derived from the obtained clusters. The unified graph is defined formally as a typed property graph

$$G = (V, E, \tau_V, \tau_E, A),$$

where V is the set of nodes, E the set of directed typed edges, $\tau_V$ and $\tau_E$ assign node and edge types, and A stores attributes including identifiers, timestamps, provenance, and confidence. The schema is aligned with STIX 2.1/TAXII 2.1 to maximize interoperability (OASIS Open, 2021a, 2021b).

Nodes represent: threat entities (ThreatActor, IntrusionSet, Campaign); TTP entities (AttackPattern, Malware, Tool); vulnerability entities (Vulnerability/CVE, Weakness/CWE); observable/indicator entities; defensive entities (DefensiveTechnique/D3FEND, Control, DetectionRule); and organization overlay entities (Asset, Account, BusinessService, NetworkZone). Edges represent identified relationships (e.g., USES_TTP, DEPLOYS, EXPLOITS, INDICATES), control mappings (MITIGATED_BY), and event propagation paths (OBSERVED_ON, HAS_VULNERABILITY).

The baseline graph contains approximately 1,184,200 vertices and 3,062,500 edges. In terms of vertex composition, Observables account for 612,400 nodes, Vulnerabilities (CVE) for 214,600, Indicators for 146,800, Organization overlay entities (Asset/Account/Service/Zone) for 179,180, and ATT&CK AttackPatterns, Malware, Tool, IntrusionSet/Campaign/ThreatActor, and D3FEND DefensiveTechniques for the remainder. The principal edge types are INDICATES (1,020,000), OBSERVED_WITH (840,000), HAS_VULNERABILITY (520,000), USES_TTP (140,000), EXPLOITS (95,000), and MITIGATED_BY (84,500). The largest connected component covers 91.2% of vertices (mean degree 5.17; 99th percentile degree 248). High-degree hubs are CVEs, common observables, and ATT&CK technique nodes — expected in CTI graphs. Figure 1 shows an example of the graph for a single attack pattern.

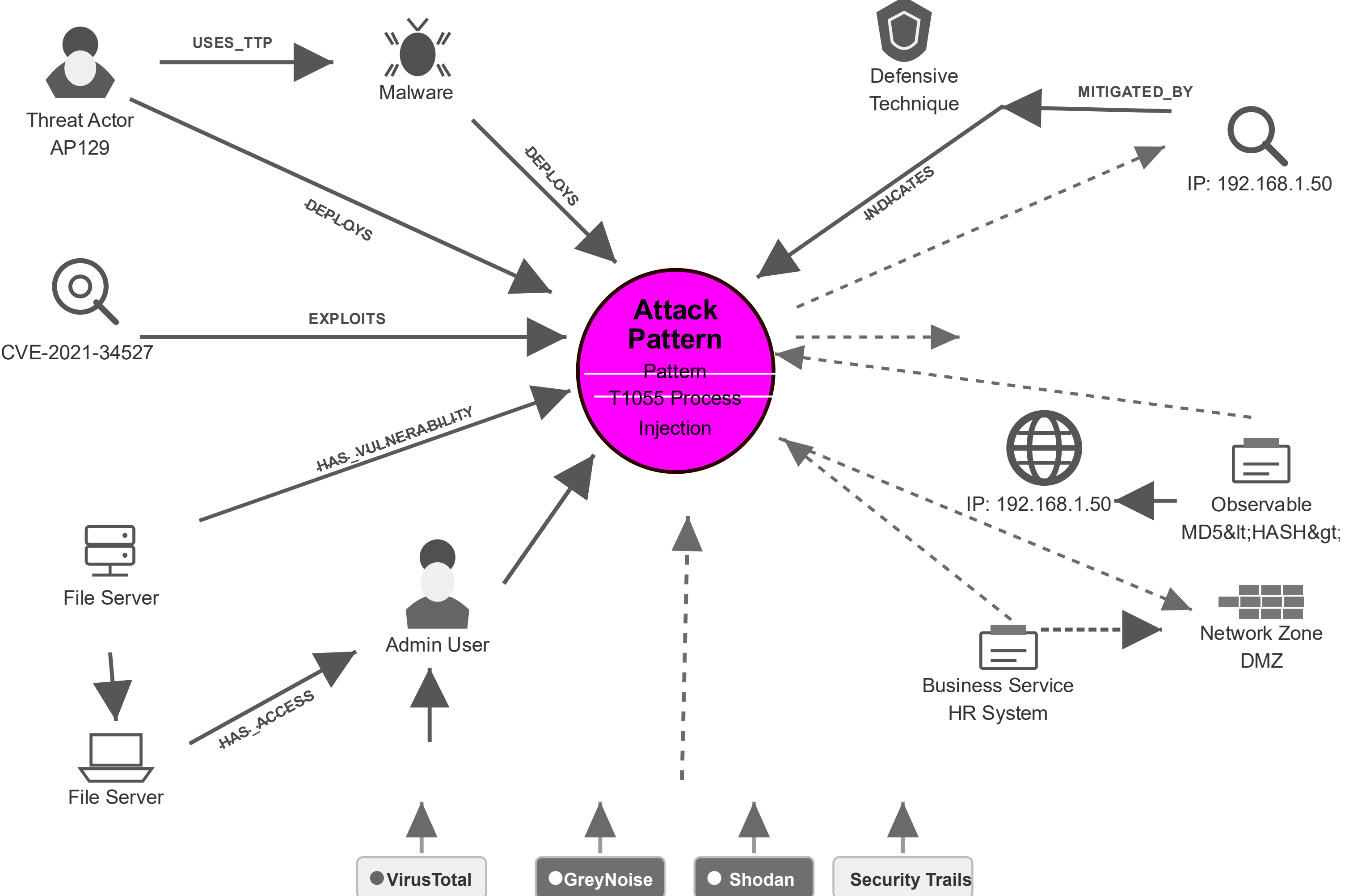

*Figure 1 – Graph extraction for a single attack pattern*

## 3.9. Metric computation and normalization

The metric registry (Artifact A3) converts the organization-specific profile into measurable variables. Each metric records its raw measure type, source, observation window, orientation, and normalization rule so that component scores remain reproducible (ISO/IEC, 2016; NIST, 2011). Table 1 presents the complete metric registry.

**Table 1.** Metrics for proposed variables

| *Variable* | *Unit* | *Gathering method/tool* | *Metric* |
|---|---|---|---|
| *Exposure* | *Technology used* | Organization assets inventory; Nmap | *# of different technologies used* |
| | *Public IPs* | *Nmap; Shodan* | *# public IPs / # necessary public IPs* |
| | *Open ports and service visibility* | Nmap; sqlmap | *# visible ports / # necessary visible ports* |

| *Variable* | *Unit* | *Gathering method/tool* | *Metric* |
|---|---|---|---|
| | *Number of Users* | *Active directory; /etc/passwd* | *# privileged users / # total users; # authenticated users / # total users; # users with shared accounts / # total users* |
| | *Number of Computers or devices* | *Nmap; Active directory* | *# registered computers / # total computers* |
| | *Asset location and access* | *Physical inventory* | *Access to assets from location / business rules requiring access from a specific location* |
| *Traceability* | *User activity registration* | *File and application logs* | *User authentication vs expected authentication* |
| | *Logs* | *OS logs; network logs; device logs; application logs; database logs* | *Automated activities vs expected transactions; devices online / expected or registered devices; traffic or transactions / expected traffic and transactions volume* |
| | *Privileges and User Permissions* | *Active directory; OS; application and system configuration* | *User authorization for activities / roles and responsibilities* |
| | *Physical monitoring and access control* | *CCTV; door locks; fire alarms; temperature sensors; presence sensors* | *Registered behaviour / expected behaviour* |
| *Motivation* | *Value of assets and information* | *Business impact analysis; asset and data classification; crown-jewel register* | *Weighted proportion of business-critical or sensitive assets among in-scope assets* |
| | *Ease of capitalization* | *Business-process mapping; backup inventory; public-disclosure review* | *Proportion of high-value assets whose compromise would enable direct extortion, fraud, service disruption, or resale* |
| | *Ease of exploitation* | *Vulnerability scanner; exposure inventory; EPSS; CISA KEV; CVSS v4* | *Proportion of exposed or reachable assets with at least one KEV-listed vulnerability or EPSS above a defined threshold* |
| | *Time required (Attack-path effort)* | *Cyber-exposure-profile graph query; attack-graph analysis; BAS/purple-team validation where available* | *Median shortest attack-path length from exposed entry points to crown-jewel assets, or proportion reachable within ≤ 2 steps* |
| | *Required attacker sophistication* | *CVSS v4 exploitability fields; BAS/pentest results* | *Proportion of relevant paths requiring only low attacker prerequisites (e.g., low AC, no AT, low PR)* |
| *Systems update* | *Patch-policy conformance* | *CMDB/asset inventory; patch-management platform; vulnerability management* | *# in-scope assets compliant with patch policy / # in-scope assets* |
| | *Test and deployment latency* | *Change-management records; patch-deployment logs* | *Median days from patch release to production deployment for standard updates* |
| | *Legacy systems with no updates (unsupported assets)* | *Asset inventory; vendor lifecycle/support data* | *# end-of-support/end-of-life assets / # in-scope assets; compensating-control coverage for such assets* |
| | *Time to update for critical updates* | *Vulnerability management; KEV/CVSS prioritization; deployment logs* | *Median or P90 days to remediate critical or KEV-listed vulnerabilities, or % remediated within SLA* |

**Table 1b.** Cluster-to-variable mapping with rationale

| **Cluster** | **Top centroid-loading terms** | **Thematic label** | **Primary variable** | **Secondary variable** | **Mapping rationale** |
|---|---|---|---|---|---|
| C1 | *public_ip, open_port, exposed_service,* | Exposed services and | Exposure | Motivation (attack-path | Dominant terms describe externally reachable conditions |

| Cluster | Top centroid-loading terms | Thematic label | Primary variable | Secondary variable | Mapping rationale |
|---|---|---|---|---|---|
| | *unregistered_device, shared_account, remote_desktop, internet_facing, vpn, cloud_exposure, external_access, unauthorized_access, privilege_escalation* | unmanaged access | | effort) | and uncontrolled access points, mapping directly to Exposure metrics (public IPs, open ports, unregistered devices, privileged/shared users). Attack-path terms partially inform Motivation sub-metric for ease of exploitation. |
| C2 | *log_gap, authentication, failed_login, unmonitored, siem, alert_fatigue, credential_misuse, session_anomaly, privilege_misuse, detection_rule, audit_trail, access_control* | Observability gaps and credential misuse | Traceability | Exposure (privileged-user concentration) | Terms cluster around logging, authentication monitoring, and detection capacity, mapping to Traceability metrics. Credential-misuse terms also feed Exposure sub-metric for privileged-user concentration. |
| C3 | *ransomware, data_theft, exfiltration, cryptocurrency, financial_data, healthcare_record, pii, extortion, payment_card, intellectual_property, command_and_control, phishing_email* | Monetizable targets and attacker incentive | Motivation | — | Terms reflect attacker goals and asset attractiveness: data types with resale or extortion value, monetization methods, and initial-access techniques targeting high-value environments. |
| C4 | *cve, unpatched, legacy_system, end_of_life, outdated_software, vulnerability_scan, patch_delay, unsupported_os, firmware, update_failure, zero_day, exploit_available* | Patching deficiency and technology debt | Systems Update | Motivation (ease of exploitation) | Terms describe patch-management failures and technology-lifecycle issues, mapping to Systems Update metrics. Exploit-availability terms secondarily inform Motivation for ease-of-exploitation. |

*Note. Top centroid-loading terms are ranked by TF–IDF weight within each cluster centroid. Primary variable assignment follows metric-registry alignment. Secondary variable is assigned when ≥ 5 of the top 30 terms directly inform metrics in a second construct.*

**Table 1c.** Exemplar incident narratives nearest to cluster centroids

| Cluster | Cosine to centroid | Exemplar incident narrative (anonymized) |
|---|---|---|
| C1 | 0.91 | *Threat actor exploited an internet-facing RDP service on an unregistered server. The device was not in the asset inventory and had default credentials. Lateral movement occurred via shared administrative accounts across the network segment.* |
| C1 | 0.88 | *External scan revealed 47 publicly exposed ports on a cloud instance provisioned outside the change-management process, including an unauthenticated API endpoint and an outdated SSH daemon.* |
| C2 | 0.93 | *Privileged account was used to exfiltrate database records over 14 days. No alerts were generated because the SIEM correlation rules did not cover the affected application, and authentication logs were not forwarded from the legacy system.* |

| Cluster | Cosine to centroid | Exemplar incident narrative (anonymized) |
|---|---|---|
| C2 | 0.87 | *Multiple failed login attempts followed by a successful authentication from an anomalous geolocation went undetected for 72 hours due to incomplete log ingestion. Audit trail reconstruction was hampered by inconsistent timestamp formats.* |
| C3 | 0.92 | *Ransomware group encrypted production databases containing financial records and patient data. Ransom demand cited the monetizable value of the data. Initial access was via a phishing email targeting accounts-payable staff.* |
| C3 | 0.86 | *Threat actor exfiltrated intellectual property and PII from a research database, subsequently offering it for sale on a dark-web marketplace. The target was selected based on publicly available information about the organization's data holdings.* |
| C4 | 0.90 | *Exploitation of CVE-2021-44228 (Log4Shell) on an unpatched application server. The vulnerability had been disclosed 87 days prior; the patch was available but had not been deployed due to testing-environment constraints and change-management backlog.* |
| C4 | 0.85 | *End-of-life operating system on an OT controller was compromised via a known firmware vulnerability. No vendor patches were available, and no compensating controls had been implemented. The device had been flagged in two prior vulnerability scans.* |

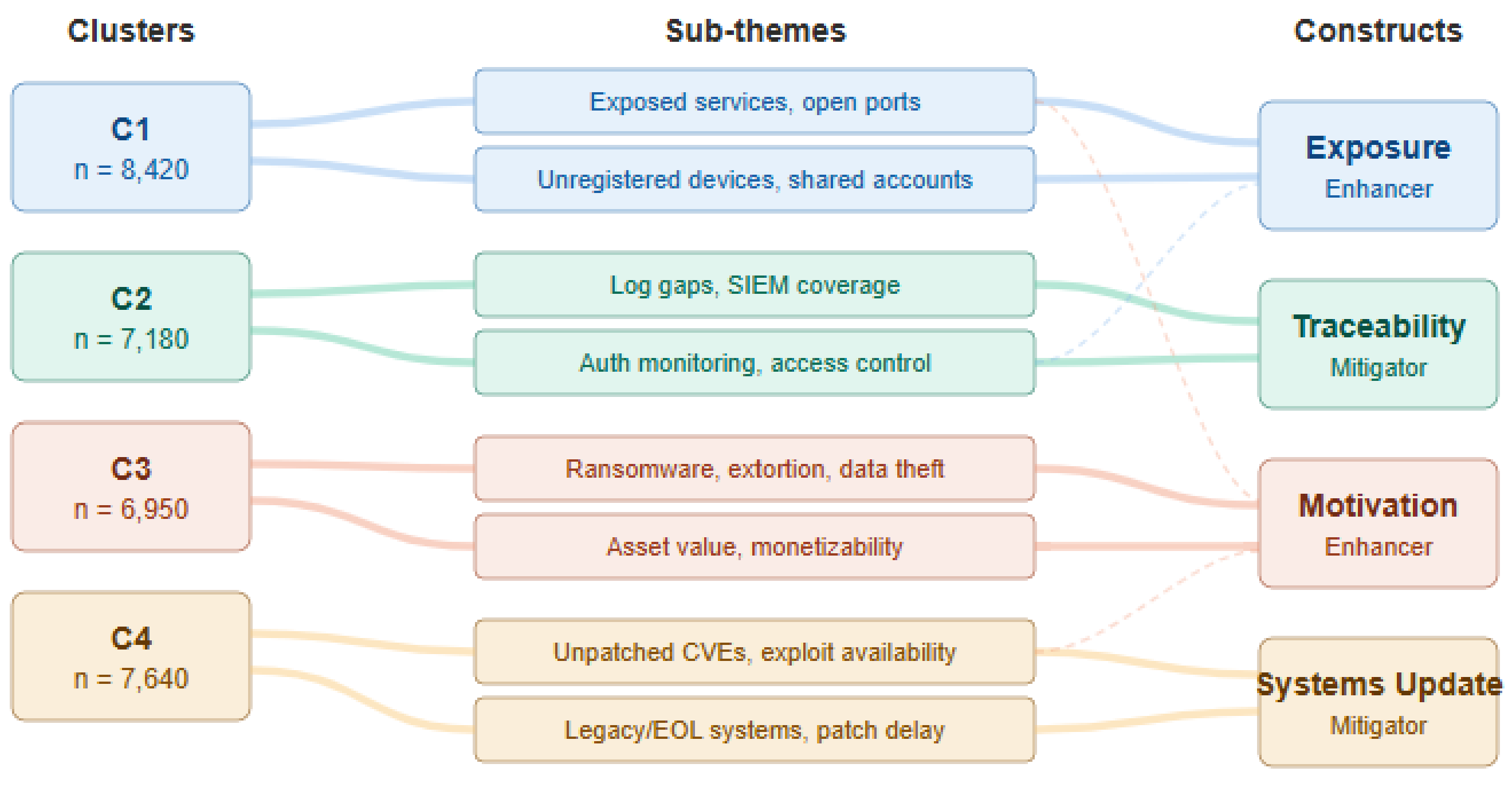

*Figure 2. Alluvial diagram for the clustering themes*

Raw metrics are normalized to the unit interval and directionally oriented. Higher Exposure and Motivation indicate worse likelihood conditions, whereas higher Traceability and Systems Update indicate stronger mitigating conditions. Where natural bounds are unavailable, the normalization range is fixed through policy thresholds or empirical study percentiles, and the choice is recorded in the computational trace.

Component scores are aggregated from their assigned metrics using a confidence-weighted mean:

$$C = \frac{\sum_{i \in \mathcal{C}} w_i \ c_i \, \widetilde{m}_i}{\sum_{i \in \mathcal{C}} w_i \ c_i},$$

where $m_i$ is the normalized metric value, $w_i$ its analytical weight, and $c_i$ a composite evidence-quality score computed from three declared dimensions. The first dimension is completeness $\kappa_i \in \{0.10, 0.25, 0.50, 0.75, 1.00\}$, defined as the fraction of a metric's required data fields populated from primary sources rather than imputed or absent. The second dimension is freshness $\phi_i \in \{0.10, 0.30, 0.55, 0.80, 1.00\}$, defined by the age of the source data relative to the observation window end-date, ranging from within-window $\phi_i = 1.00$ to undated or older than 180 days $\phi_i = 0.10$. The third dimension is source authority $s_i \in \{0.20, 0.50, 0.75, 1.00\}$, defined by the relationship between the data source and the construct being measured: canonical primary sources such as a CMDB for asset inventory or Active Directory for account data receive $s_i = 1.00$; secondary derived sources such as network-scan-inferred asset lists receive $s_i = 0.75$; tertiary or analyst-estimated sources receive $s_i = 0.50$; and unverified or assumed values receive $s_i = 0.20$. The composite score $c_i = \kappa_i \times \phi_i \times s_i$ ensures that weakness in any single dimension degrades overall confidence without being offset by strength in another — a complete and authoritative source that is six months stale is appropriately down-weighted. The scoring bands for all three dimensions are recorded in the metric registry (Artifact A3) alongside each metric's raw value and normalization trace, ensuring that every $c_i$ is recoverable from declared source metadata rather than from undocumented practitioner judgement. Two conservative rules apply irrespective of $c_i$: uncertainty in a mitigating metric cannot improve the aggregate score, so missing logging coverage does not increase Traceability and absent patch records do not improve Systems Update.

In the default configuration, metric weights are equal within each construct unless a justified alternative is reported. Figure 3 presents the cyber-exposure profile and its attributes.

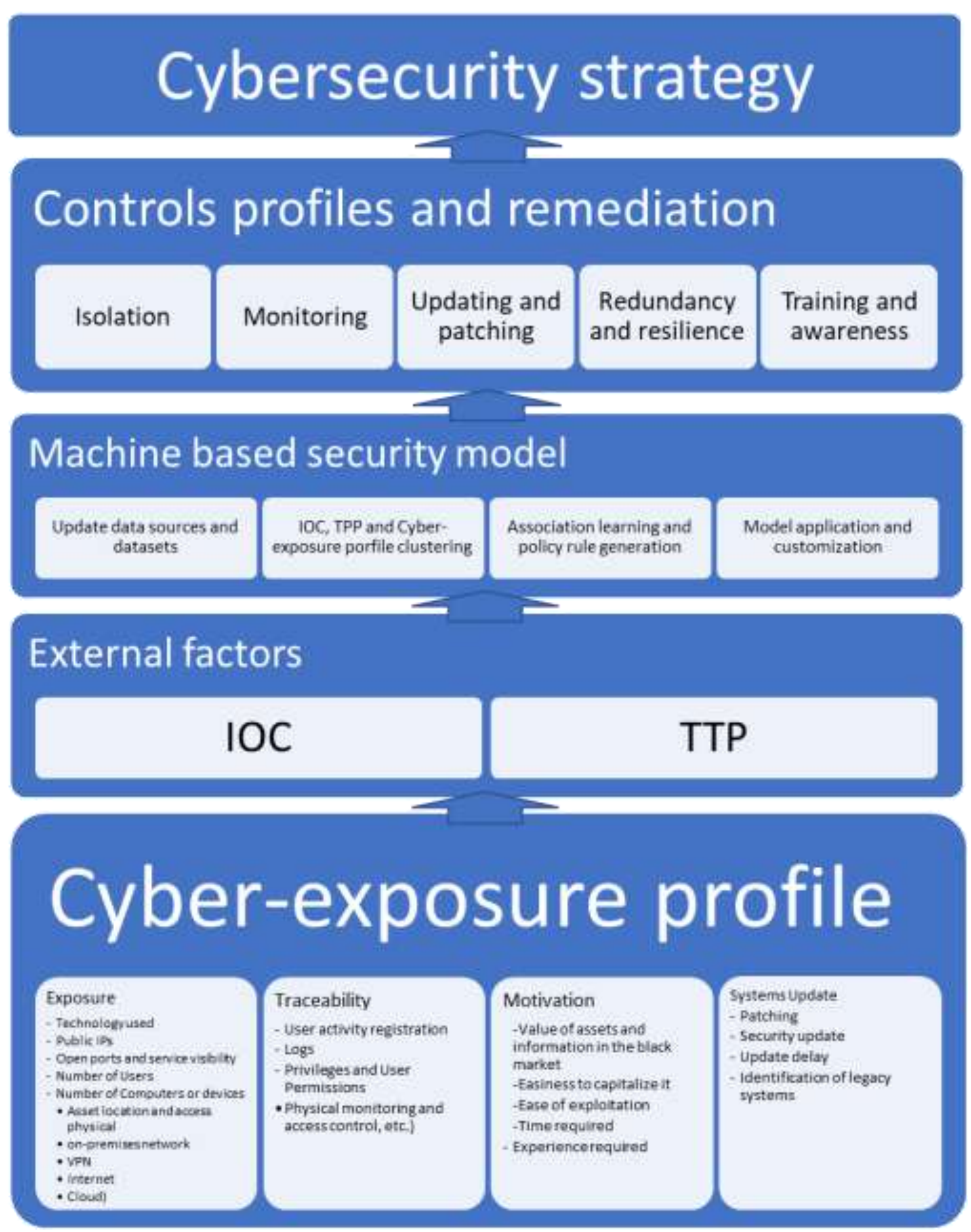

*Figure 3 – Cybersecurity strategy data model and cyber-exposure profile*

### 3.10. Likelihood computation and calibration

The likelihood computation rule (Artifact A4) maps the four normalized component scores — Exposure (E), Traceability (T), Motivation (M), and Systems Update (U) — to a bounded indicator L. The mapping satisfies four design requirements: monotonicity (the score increases when E or M increases and decreases when T or U improves); boundedness (the output remains in (0,1)); robustness to normalization choices; and auditability (each component's contribution can be inspected transparently).

A log-additive/logistic specification is adopted as the reference form. Its main advantages are that it satisfies all four requirements while preserving an interpretable additive structure in the latent space (Agresti, 2013; McCullagh & Nelder, 1989; OECD & JRC, 2008). Let E, T, M, U ∈ [0,1] denote the normalized component scores, and let $\varepsilon > 0$ be a small positive constant to avoid singularities. The latent score is:

$$z = \alpha \ln(\varepsilon + E) + \beta \ln(\varepsilon + M) - \gamma \ln(\varepsilon + T) - \delta \ln(\varepsilon + U)$$
$$\alpha, \beta, \gamma, \delta \geq 0$$

The bounded incident-likelihood indicator is then:

$$L = \frac{1}{1 + e^{-z}}$$

Higher values of Exposure and Motivation increase the latent score, whereas higher Traceability and Systems Update decrease it. The logistic transformation ensures $L \in (0,1)$, making the score suitable for monitoring and longitudinal comparison.

An equivalent way to view the same model is through the odds form. Because the logistic mapping implies $\frac{L}{1-L} = e^z$, the odds of incident likelihood can be written as:

$$\frac{L}{1-L} = \frac{(\varepsilon + E)^\alpha (\varepsilon + M)^\beta}{(\varepsilon + T)^\gamma (\varepsilon + U)^\delta}.$$

## 3.11. Control prioritization output

Artifact A5 consumes the four component scores and the cyber-exposure profile graph to produce a ranked, graph-traceable control list. It operates in four sequential steps.

Step 1 — Component leverage ranking. The partial derivatives of the logistic function with respect to each component score identify which dimension has the greatest marginal influence on L at its current value. Because L(1−L) is a common factor, the signed marginal leverage for each component simplifies to:

$$\lambda_E = \frac{\alpha}{\varepsilon + E}, \lambda_M = \frac{\beta}{\varepsilon + M} \text{(enhancers: positive leverage)}$$
$$\lambda_T = \frac{\gamma}{\varepsilon + T}, \lambda_U = \frac{\delta}{\varepsilon + U} \text{(mitigators: opportunity leverage)}$$

For the enhancers E and M, a high $\lambda$ indicates that the component is already near its upper bound and difficult to reduce further through a single control cycle; a low $\lambda$ indicates greater sensitivity to incremental change. For the mitigating components $T$and $U$, a high value of $\lambda_T$or $\lambda_U$indicates that the component remains well below its target level, so that even small improvements can produce comparatively large reductions in $L$. Components are ranked according to their improvement opportunity. For mitigators, this is defined as

$$O_C = \lambda_C\ (C_{\text{target}} - C), C \in \{T, U\},$$

whereas for enhancers it is defined as

$$O_C = \lambda_C\ (C - C_{\text{target}}), C \in \{E, M\}.$$

Here, $C_{\text{target}}$denotes the policy threshold or empirical percentile recorded in Artifact A3 for the corresponding component. The component with the highest opportunity score is designated the primary control target for the current observation window.

Step 2 — Metric-level deficit identification. Within the primary target component, individual metrics are ranked according to their normalized deficit. For mitigating metrics, the deficit is defined as

$$\text{deficit}(m_i) = m_{i,\text{target}} - m_i,$$

whereas for enhancing metrics it is defined as

$$\text{deficit}(m_i) = m_i - m_{i,\text{target}}.$$

Metrics whose deficit exceeds a declared threshold are flagged as high-priority inputs for the graph query in Step 3. Unless otherwise specified, the default threshold is 0.20 in normalized units. This threshold is recorded in Artifact A3 and may be adjusted by an analyst, provided that the adjustment is explicitly justified.

Step 3 — Graph-based control retrieval. For each high-deficit metric, the cyber-exposure profile G~o,w~ is queried to retrieve applicable defensive controls. The query traverses from the organization's affected assets through the typed-edge structure of the unified graph: for Exposure deficits, the path is Asset → (via HOSTS or HAS_VULNERABILITY) → Vulnerability or AttackPattern → (via MITIGATED_BY) → DefensiveTechnique → Control; for Traceability deficits, the path traverses LogSource or DetectionRule nodes with the identified gap to D3FEND detection and hardening techniques; for Systems Update deficits, the path traverses Vulnerability nodes with active EXPLOITS edges and KEV listing or EPSS above threshold to patch and configuration Controls. All retrieved Controls are D3FEND-aligned and carry the MITIGATED_BY provenance chain from the source condition, preserving full traceability from the metric deficit to the recommended action.

Step 4 — Control scoring and ranking. Each candidate control c is scored as:

$$\text{score}(c) = \text{coverage}(c) \times \text{severity}(c) \times [1 + 0.25 \times \text{cross}(c)]$$

where coverage(c) is the number of high-deficit metrics addressed by c normalized to [0,1]; severity(c) is the mean CVSS base score or asset-criticality weight of the affected conditions, normalized to [0,1]; and cross(c) is the count of additional components beyond the primary that c also improves, rewarding controls with multi-dimensional impact. Controls are output in descending score order. Each entry in the A5 output records: control identifier, D3FEND and ATT&CK alignment, targeted component(s), metrics addressed, the improvement opportunity score for those metrics, and the expected reduction in L if each addressed metric moves from its current value to its policy threshold.

## 3.12. Graph querying and instantiation

Artifact A2 is the instantiation of the graph based on the organization's telemetry and context. The empirical unit of analysis is the organization–time window pair: for each organization, a fixed observation window is used over which evidence is collected, normalized, and aggregated into the cyber-exposure profile. This unit is necessary because cyber conditions are dynamic — asset inventories change, systems are updated, new exposures emerge, and attacker incentives shift.

Based on the first-layer graph (containing AttackPattern, Malware, Tool, Vulnerability, Weakness, Indicator, Control, and Incident entities), a second layer is built from organization-specific information: Asset, Service, Account, LogSource, DetectionRule, BusinessProcess, Location, and ObservationWindow. Formally, for organization o and observation window w = [$t_s$, $t_e$], the cyber-exposure profile is the induced subgraph

$G_{o,w} = (V_{o,w}, E_{o,w}),$

where the seed set is

$$S_{o,w} = \{v \in V: v \text{ is directly owned, operated, or declared by } o \text{ and overlaps } w\}.$$

The node set of the profile is then

$$V_{o,w} = \{v \in V: v \in S_{o,w} \ \vee \ \exists s \in S_{o,w}, \exists p = (s \rightsquigarrow v), \text{type}(p) \in R^*_{adm}, \text{and } overlap_t(p,w) = 1\},$$

where $R^*_{adm}$ denotes admissible relation paths used by the framework, such as ownership, hosting, execution, vulnerability association, observed behavior, detection linkage, business dependency, and control mapping. The edge set is

$$E_{o,w} = \{e = (u, r, v) \in E : u, v \in V_{o,w} \wedge overlap_t(e, w) = 1\}.$$

The seed set includes the organization's assets, services, accounts, business processes, and declared observation window, and additional nodes are included when reachable via admissible relation paths (ownership, hosting, execution, vulnerability association, observed behavior, detection linkage, business dependency, control mapping) that overlap the window. Edges are retained when both endpoints are in $V_{o,w}$ and the edge overlaps the window.

In operational terms, the profile therefore contains: (i) organization-local entities and their states within the window; (ii) directly connected CTI entities relevant to the deployed technologies, observed exposures, or incidents; and (iii) the typed relations needed to compute the metrics in Table 1 deterministically.

An example extraction query in Cypher:

```
MATCH (org:Organization {id: $org_id})-[:OWNS|OPERATES|HAS_PROCESS]->(s)
WHERE coalesce(s.valid_from, date('1900-01-01')) <= $window_end
  AND coalesce(s.valid_to, date('9999-12-31')) >= $window_start
OPTIONAL MATCH p=(s)-[:HOSTS|RUNS|HAS_ACCOUNT|HAS_VULNERABILITY|
                     OBSERVED_IN|USES_TTP|DETECTED_BY|
                     AFFECTS|DEPENDS_ON|MITIGATES*0..3]-(n)
WHERE coalesce(n.valid_from, date('1900-01-01')) <= $window_end
  AND coalesce(n.valid_to, date('9999-12-31')) >= $window_start
  AND (n.organization_id = $org_id OR n:Vulnerability OR n:Weakness
       OR n:AttackPattern OR n:Indicator OR n:Control OR n:Incident)
WITH collect(DISTINCT s) + collect(DISTINCT n) AS nodes
MATCH (u)-[r]->(v)
WHERE u IN nodes AND v IN nodes
  AND coalesce(r.valid_from, date('1900-01-01')) <= $window_end
  AND coalesce(r.valid_to, date('9999-12-31')) >= $window_start
RETURN nodes, collect(DISTINCT r) AS edges
```

Conflicts between external CTI and internal telemetry are handled explicitly through three rules. First, entities are merged only when canonical identifiers exist (CVE IDs, ATT&CK technique IDs, approved asset keys); absent such identifiers, ambiguous entities are retained as distinct nodes with a provisional equivalence annotation pending review. Second, source precedence is attribute-specific: internal telemetry is authoritative for organization-local state variables (whether an asset exists, whether a service is currently exposed, whether a patch is installed, whether logging is enabled, whether an account is privileged); external CTI is authoritative for global threat semantics (vulnerability definitions, exploit reports, ATT&CK technique meaning, weakness classification, control mappings). Third, when conflicting assertions remain within the same attribute class, the framework applies recency-plus-authority ordering for computation while preserving both assertions in the provenance layer.

Two conservative scoring rules apply throughout: uncertainty in a mitigating attribute cannot improve the score — missing or disputed logging evidence does not increase Traceability, and disputed patch evidence does not improve Systems Update; and CTI relevance is not confused with local exposure status — when internal evidence indicates a vulnerability has been remediated, the CTI relation is preserved as contextual knowledge while the local exploitable-state metric is computed from current telemetry. This separation prevents CTI relevance from being confused with local exposure status and ensures that the cyber-exposure profile reflects the organization's actual current condition rather than the aggregate threat landscape.

Confidence is paired with each metric when evidence quality differs across sources, tempering the contribution of weaker or stale measurements without defining an additional risk construct. Figure 4 shows the cyber-exposure profile extraction from the unified graph.

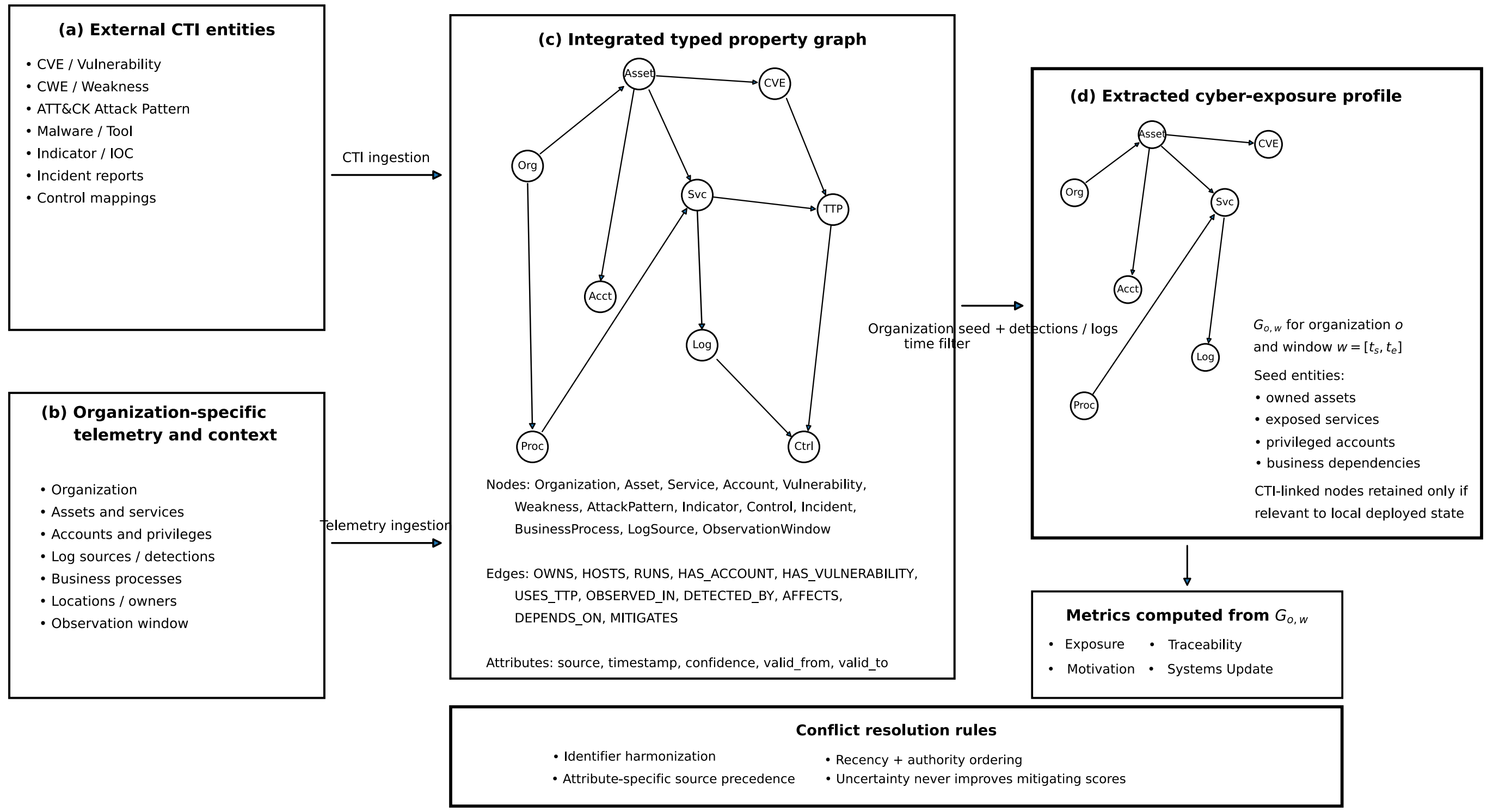

*Figure 4 – cyber-exposure profile extraction from the unified graph*

*Note. The unified graph integrates external cyber threat intelligence with organization-specific telemetry and business context. The cyber-exposure profile is the organization and time-bounded induced subgraph used to compute the four likelihood dimensions.*

## 3.13. Framework

In line with Peng (2011), the framework is presented as an explicit protocol with configuration capture, versioning, provenance, and preservation of intermediate artifacts:

- **C1** – Define scope and unit of analysis: organizational context, evaluation time window(s), and event definition.

- **C2** – Assemble data sources: external cyber knowledge and event datasets, plus organization telemetry and asset context.
- **C3** – Map sources to the unified data model: transform each source into the entity–relation schema (A1).
- **C4** – Integrate into a graph instance: create the integrated knowledge graph and validate structural consistency.
- **C5** – Instantiate the organization profile: extract the organization-constrained subgraph to form the cyber-exposure profile (A2).
- **C6** – Compute and normalize metrics: generate the metric vector from A2 using A3, normalized to [0,1].
- **C7** – Compute likelihood and produce recommendations: apply the likelihood computation rule (A4) and generate control prioritization outputs (A5).

# 4. Experimental setup and case-study design

## 4.1. Case selection and anonymization

The empirical evaluation used a multiple-case study to examine whether the proposed framework can be instantiated consistently across organizations with different infrastructures, exposure profiles, and defensive baselines, while preserving traceability from evidence to outputs (Runeson & Höst, 2009; Yin, 2018). The study covered 15 organizations in five sectors: finance (n=4), healthcare (n=3), manufacturing (n=4), higher education (n=2), and public administration (n=2). Cases were included only if they provided sufficient telemetry for Table 1 metrics, supported longitudinal observation, and had a measurable implemented or planned security intervention. All organizations were anonymized as O1–O15.

## 4.2. Observation horizon and windowing

The evaluation covered a longitudinal horizon from August 2024 to August 2025, enabling repeated measurement of the cyber-exposure profile, component scores, and likelihood indicator under changing operational conditions. The default design used fixed monthly windows (13 per organization), balancing temporal sensitivity with measurement stability and aligning with common security and compliance reporting cycles. For each organization and window w, only time-bounded observations were used; when a control change occurred, an intervention time $\tau_0$ was recorded to divide windows into pre- and post-intervention segments.

## 4.3. Outcome definitions

Outcomes were derived from incident records and security reports using a common framework across cases. An incident was counted only when it crossed the organization's incident-handling threshold and required formal containment, eradication, coordinated response, or equivalent action. Routine alerts or benign anomalies were excluded. The primary outcome was the window-level incident count $I_{(o,w)}$, summarized through pre/post mean window counts relative to intervention time $\tau_o$:

$$\text{Incident Reduction}_o = \frac{\bar{I}_o^{pre} - \bar{I}_o^{post}}{\bar{I}_o^{pre}} \times 100.$$

The detection-and-response interval is defined as the elapsed time from the first recorded alert or ticket opening to formal incident containment or closure. Where organizations used different ticketing tools, timestamps were harmonized to a common epoch at ingestion; cases with absent or unreliable closure timestamps were excluded rather than imputed.

## 4.4. Interventions

Interventions were defined as deliberate, documented changes to the organizational control environment introduced to improve the cyber-exposure profile and reduce the computed likelihood indicator. Each intervention was treated as a time-stamped event set with organization-specific implementation time $\tau_o$. The intervention record for each organization included: (a) intervention time $\tau_o$; (b) affected control categories; and (c) implementation evidence (configuration snapshots, policy updates, tooling deployment dates, audit-log traces).

Interventions were grouped by framework dimension. Exposure-oriented interventions included registering unmanaged devices, reducing unnecessary public exposure, segmenting exposed services, hardening internet-facing systems, and reducing shared or privileged accounts. Traceability-oriented interventions included expanding logging coverage across endpoints and network devices, improving SIEM correlation rules, strengthening detection logic for privileged-account activity, and ensuring that authentication events from all critical systems were forwarded for correlation. Systems Update interventions addressed patching cadence, legacy system replacement or compensating-control coverage, and critical remediation SLA enforcement. More broadly, organizations documented their current context, established a baseline cyber-exposure profile, selected controls through model outputs, implemented them, and continued monitoring incidents and vulnerabilities to assess effectiveness. Because intervention packages were often composite, the evaluation does not attribute all outcome variation to a single control; post-intervention changes are interpreted relative to the documented intervention set for each window.

## 4.5. Parameter settings and directional consistency

All cases used a common parameterization strategy so that comparisons reflected measured cyber condition rather than organization-specific tuning. Raw measurements were computed according to Table 1, normalized to the unit interval, and aggregated using the procedure described above. The bounded likelihood score was computed with the log-additive formulation using a fixed stability constant $\varepsilon = 10^{-6}$.

The reference parameterization ($\alpha = \beta = \gamma = \delta = 1$) is justified on two grounds. First, equal weighting constitutes the recommended neutral prior in composite indicator construction when empirical calibration data are insufficient (OECD & JRC, 2008). Second, setting all coefficients to unity enforces symmetry between the two likelihood-enhancing components and the two mitigating components — a conservative and directionally appropriate default that defers asymmetric claims to future calibration. An internal robustness check comparing the reference log-additive specification with a simple linear alternative yielded Spearman rank correlation $r_s = .97$ across the 15 baseline profiles, indicating that the choice between functional forms has negligible effect on the relative ranking of organizations at the current level of construct-score variation.

Sensitivity analysis across low (0.5), reference (1), and high (2) uniform scenarios, as well as two asymmetric scenarios, showed that the mean incident reduction varied by no more than 3.2 percentage points and the pre-intervention directional consistency Pearson correlation remained within [.696, .741] across the tested range. Empirical calibration of these coefficients through maximum-likelihood estimation or Bayesian updating against a larger cross-organizational outcome dataset remains an explicit priority for future work.

The directional consistency check tests whether the full pre-intervention likelihood score L, computed from baseline data before any control changes, is positively associated with the directly observed pre-intervention qualifying incident count. This test uses only pre-intervention data and is independent of the framework's subsequent control recommendations. One detection-related limitation applies: organizations with higher Traceability are expected to detect and record more incidents per unit of true exposure, which attenuates the correlation between L and raw incident count in proportion to between-organization variation in T. The check therefore constitutes directional evidence rather than a bias-free predictive test.

## 4.6. Threats to validity

As an observational, multiple-case longitudinal evaluation, this study faces validity threats typical of case-based and quasi-experimental pre/post research (Runeson & Höst, 2009; Shadish et al., 2002; Yin, 2018).

A key risk is that improvements in the output score are guaranteed by design whenever recommended actions are taken, because those actions directly alter measured inputs. To address this, the directional consistency check uses only pre-intervention data to test whether the framework orders organizations consistently with their observed incident frequencies before any control changes were introduced. The resulting associations (Pearson $r = .736$, $p = .002$; Spearman $r_s = .720$, $p = .003$) provide directional but not causal support. Full causal identification would require a comparison condition — organizations that did not receive framework-guided interventions — recommended for future work via a stepped-wedge design.

Confounding remains possible because outcome changes may coincide with other organizational changes; this is mitigated through explicit intervention records, within-organization pre/post comparison, and triangulation with timing and observability outcomes. Reporting bias and data heterogeneity are relevant because incident thresholds, telemetry coverage, and SOC maturity differ across organizations. The framework addresses these through a common incident definition, threshold-based counting, and explicit handling of traceability, confidence, and source availability. External validity is limited by purposive case selection, supporting analytic rather than statistical generalization. Future evaluations should include matched comparison organizations or a stepped-wedge design to provide a stronger basis for attributing observed gains to the framework's distinctive features rather than to the general benefit of any structured security engagement.

### 4.7. Case instantiation procedure

Each case was instantiated through a standardized procedure consistent with case-study guidance (Runeson & Höst, 2009; Yin, 2018): define organizational scope and observation window; map evidence into the unified data model; extract the organization-specific cyber-exposure profile for each time window; compute metrics and likelihood scores; record and align documented control changes; and evaluate pre/post outcomes. Interventions were recorded as dated control sets, allowing comparisons to be tied to documented changes rather than undocumented drift.

## 5. Results

### 5.1. Descriptive results

Across the 15 evaluated organizations, the most consistent baseline pattern was the coexistence of substantial unmanaged-device exposure, concentrated privileged access, and low or inconsistent logging coverage, observed across all five sectors. After implementation of the control sets derived from the framework, all organizations showed improvements in registered-device coverage, reductions in privileged-user concentration, and marked gains in logging coverage, with post-intervention logging levels exceeding 80% in every case.

The central operational outcome was a reduction in incident frequency after intervention. At the sample level, mean incident reduction was 35% (SD = 18.0%), and the paired pre/post comparison was statistically significant, $t(14) = 5.78$, $p < .001$. Detection-and-response performance also improved, with a mean reduction of 23.5% (SD = 9.01%); the sample median fell from 19 h (IQR = 16–23 h) before intervention to 14 h (IQR = 11–17 h) afterward. These results indicate that the framework is associated not only with fewer incidents, but also with faster operational handling of those incidents that still occur.

The strongest incident-reduction results were observed in public administration (51% and 47%) and two manufacturing cases (50% and 43%), whereas finance and higher-education cases showed more moderate but still material gains. Response-time improvement was more heterogeneous than incident reduction: most organizations improved, but one finance case (O3) showed approximately no measurable reduction in the detection-and-response interval. The comparatively low reductions in finance (O1–O4: 4–9%) most plausibly reflect stronger pre-intervention baseline postures — organizations operating under CNBV, PCI-DSS and ISO 27001 obligations typically deploy denser monitoring and stricter access controls before any engagement, leaving a narrower absolute margin for improvement. This means the percentage-based reduction metric may understate the security gain achieved in already-mature organizations. The flat detection-and-response result for O3 is consistent with a separate constraint: SOC capacity and correlation logic were not materially altered during the intervention window for that case, so exposure-layer improvements did not translate into faster operational handling. Both patterns reinforce the framework’s diagnostic value for identifying which dimension — exposure or traceability — remains the binding constraint for a given organization.

**Table 2.** Implementation evaluation across the 15 organizations.

| Org. | Sector | Total devices (baseline) | Unregistered (baseline) | Privileged users (baseline) | Logging (baseline) | Registered devices (post) | Privileged users (post) | Logging (post) | Qualifying incidents (pre, 6 months) | Qualifying incidents (post, 6 months) | Incident reduction | Detection & response Δ | Compliance uplift |
|---|---|---|---|---|---|---|---|---|---|---|---|---|---|
| O1 | Finance | 275 | 86 | 27 | <85% | 200 | 9 | >90% | 11 | 10 | 9% | −26% | PCI-DSS pass |
| O2 | Finance | 361 | 134 | 19 | <76% | 250 | 11 | >90% | 14 | 13 | 7% | −11% | ISO 27001 gap↓60% |
| O3 | Finance | 437 | 369 | 45 | <91% | 400 | 16 | >93% | 20 | 19 | 5% | ≈0% | n/a |
| O4 | Finance | 198 | 72 | 11 | <79% | 164 | 7 | >85% | 24 | 23 | 4% | −18% | ISO 27001 rdy |
| O5 | Health. | 612 | 215 | 52 | <25% | 560 | 18 | >92% | 20 | 11 | 45% | −29% | HIPAA pass |
| O6 | Health. | 458 | 172 | 31 | <40% | 420 | 12 | >90% | 35 | 18 | 49% | −31% | ISO 27799 rdy |
| O7 | Health. | 233 | 55 | 17 | <45% | 220 | 6 | >88% | 25 | 14 | 44% | −25% | HIPAA gap↓55% |
| O8 | Manufact. | 795 | 301 | 60 | <20% | 730 | 21 | >93% | 32 | 20 | 38% | −23% | IEC 62443 stage 1 |
| O9 | Manufact. | 1083 | 476 | 88 | <15% | 1020 | 25 | >95% | 30 | 15 | 50% | −33% | IEC 62443 stage 1 |
| O10 | Manufact. | 504 | 118 | 24 | <35% | 465 | 10 | >90% | 30 | 17 | 43% | −21% | ISO/IEC 27001 rdy |
| O11 | Manufact. | 369 | 97 | 21 | <40% | 340 | 9 | >86% | 25 | 15 | 40% | −19% | CMMC L2 rdy |
| O12 | Higher Ed | 917 | 287 | 48 | <30% | 871 | 17 | >88% | 35 | 19 | 46% | −27% | MAAGTICSI gap↓50% |
| O13 | Higher Ed | 364 | 69 | 14 | <25% | 331 | 8 | >90% | 26 | 15 | 42% | −24% | MAAGTICSI rdy |
| O14 | Public Adm. | 1126 | 413 | 71 | <10% | 1040 | 24 | >94% | 49 | 24 | 51% | −35% | MAAGTICSI base |
| O15 | Public Adm. | 808 | 266 | 46 | <20% | 756 | 15 | >92% | 30 | 16 | 47% | −30% | MAAGTICSI base |

*Note. Logging coverage is reported as coarse deployment bands because collection precision varied across operational environments; reduction values are reported relative to each organization's pre-intervention baseline. O3 has a small, well-monitored server estate and a large shadow population of BYOD or IoT devices not enrolled in any MDM. MAAGTICSI is Mexico-specific federal standard for public administration.*

**Table 3.** Summary quantitative results.

| Statistic | Value | Additional information |
|---|---|---|
| Mean incident reduction | 35% | SD = 18% |
| Mean detection-and-response-time reduction | 23.5% | SD = 9.01% |
| Correlation: change in Exposure metric vs. incident reduction | r = .751, 95% CI [.39, .91] | t(13) = 4.10, p = .001 |
| Correlation: change in Traceability metric vs. response-time reduction | r = .700, 95% CI [.29, .89] | t(13) = 3.54, p = .004 |
| Paired-samples test: pre/post incident counts | t(14) = 5.78 | p < .001 |
| Median detection-and-response interval | 19 h → 14 h | IQR 16–23 h → 11–17 h |

*Note. The reported t and r statistics are computed from observed pre/post outcomes and component-score changes (ΔE, ΔT); they are not based on post hoc coefficient fitting of the bounded likelihood score.*

## 5.2. Model traceability

For each organization and time window, the analysis follows a computation chain connecting: (a) the unified graph representation of cyber-risk evidence, (b) the organization-specific cyber-exposure profile, (c) metric computation and normalization, (d) bounded likelihood scoring, and (e) observed operational outcomes. At the sample level, this traceability is reflected in two informative associations. The change in the Exposure component score (ΔE) was positively associated with incident reduction, r = .751, 95% CI [.39, .91], t(13) = 4.10, p = .001. The change in the Traceability component score (ΔT) was positively associated with reduction in the detection-and-response interval, r = .700, 95% CI [.29, .89], t(13) = 3.54, p = .004. Both confidence intervals lie entirely above zero, confirming the direction of the associations. Although these associations do not establish full causal identification, they are directionally consistent with the framework's structure: reductions in unmanaged exposure are associated with fewer incidents, whereas improvements in observability are associated with faster operational handling.

The directional consistency check examined whether the full pre-intervention likelihood score L, computed from all four Table 1 components under the reference parameterization, is positively associated with the directly observed pre-intervention qualifying incident count. Across the full sample, the baseline L score (mean = 0.667, SD = 0.181, range: 0.367–0.891) was positively and significantly associated with the pre-intervention qualifying incident count: Pearson r = .736, 95% CI [.36, .91], t(13) = 3.92, p = .002; Spearman rs = .720, t(13) = 3.74, p = .003.

A simple regression yielded a slope of 37.8 incidents per unit of L (SE = 9.6), $R^2$ = .542. No organization produces a standardized residual below −1.5 or above 2.1.

O14 (Public Administration) is the most notable case at +2.09: its L score of 0.887 — the second highest in the sample — already predicts a high incident count, but the observed 49 incidents in six months exceeds the regression prediction of 35.4. This is interpretable as a ceiling effect of the bounded score: extreme values of E and M compress the marginal contribution of additional exposure into an increasingly narrow slice of the (0,1) interval, while incident counts continue to accumulate linearly outside that bound.

That O9 (Manufacturing), with the *highest* baseline L in the sample (0.891), shows a *negative* standardized residual (−0.85; 30 observed incidents versus 35.5 predicted) is consistent with this interpretation rather than contradictory to it. The ceiling effect compresses L scores for both organizations to within 0.004 of each other, rendering the score unable to discriminate between them at this level of granularity. But compression of the score does not prevent the organizations' true incident frequencies from diverging along a dimension the score does not capture. Crucially, the Traceability difference between O9 and O14 (T = 0.288 versus T = 0.274) does not explain O9's lower count: O9's marginally higher logging coverage means the detection funnel would be expected to surface *more* incidents in O9 relative to true exposure, not fewer. The divergence is better attributed to sector-specific attack frequency. Manufacturing environments matching O9's profile face a threat landscape dominated by targeted, lower-frequency intrusions — adversaries expending significant effort per campaign. Public-administration entities such as O14 attract a wider variety of attacker types simultaneously: opportunistic ransomware campaigns, politically motivated actors, and high-volume automated scanning of citizen-facing services. This breadth of attacker population drives incident volume beyond what asset value and exploitability alone — the basis of the current Motivation operationalization — would predict. The ceiling effect and the sector-frequency gap therefore produce opposite residual signs from nearly identical L values: a model behaviour that itself illustrates both the utility and the current limits of the Motivation dimension, and that points directly to the deeper incorporation of sectoral and regulatory context recommended in Future Research.

The sector-level ordering is directionally consistent throughout. Finance organizations (mean L = 0.419, mean pre-intervention count = 17.3) show the lowest values on both dimensions, consistent with stronger pre-intervention compliance and monitoring baselines. Healthcare (mean L = 0.653, mean count = 26.7) and Manufacturing (mean L = 0.799, mean count = 29.3) are intermediate. Public Administration (mean L = 0.871, mean count = 39.5) shows the highest values on both. Higher Education (mean L = 0.715, mean count = 30.5) departs slightly from the monotone sector ordering relative to its incident count, consistent with the framework's expectation that open-network architectures and large transient user populations generate incident volumes that the current Motivation operationalization may partially underweight.

Three constraints bound what the check establishes: the wide confidence intervals at N = 15, the detection-funnel attenuation from Traceability variation (organizations with higher Traceability detect and record more incidents per unit of true exposure, providing a conservative lower-bound estimate), and the cross-sectional nature of the check at organization level. Results should be read as directional evidence rather than formal predictive validation.

## 5.3. Worked example using one real organization

Organization O14, a public-administration case, showed one of the weakest baseline postures in observability and unmanaged exposure, and one of the strongest post-intervention gains. At baseline, it had 1,126 devices, 413 unmanaged; 71 users with privileged access; and logging coverage below 10%. After intervention, registered devices increased to 1,040, privileged users fell to 24, and logging coverage exceeded 94%. Outcomes included a 51% reduction in incidents, a 35% reduction in detection-and-response time, and MAAGTICSI-baseline compliance uplift. Greater device registration and lower privileged-access concentration indicate lower Exposure, while expanded logging indicates higher Traceability — both consistent with the framework's interpretation.

**Table 4.** Worked example: O14 baseline and post-intervention profile.

| Indicator | Baseline | Post-intervention | Interpretation |
|---|---|---|---|
| Registered devices | 713 of 1,126 (63.3%) | 1,040 of 1,126 (92.4%) | Substantially reduced unmanaged-asset exposure. |
| Privileged users | 71 | 24 | Reduced concentration of high-impact accounts. |
| Devices with logging | <10% | >94% | Large gain in observability and traceability. |
| Incident reduction | — | 51% | Largest observed reduction in the sample. |
| Detection and response interval | — | −35% | Fastest timing improvement in the sample. |
| Compliance uplift | — | MAAGTICSI baseline | Improved governance and assurance posture. |

**Table 5.** Baseline and post-intervention component scores and bounded likelihood indicator for all 15 organizations.

| Org | Sector | Phase | E | T | M | U | L |
|---|---|---|---|---|---|---|---|
| O1 | Finance | Baseline | 0.497 | 0.709 | 0.604 | 0.659 | 0.391 |
| O1 | Finance | Post | 0.327 | 0.879 | 0.529 | 0.821 | 0.193 |
| O2 | Finance | Baseline | 0.442 | 0.669 | 0.597 | 0.633 | 0.384 |
| O2 | Finance | Post | 0.327 | 0.851 | 0.528 | 0.787 | 0.205 |

| Org | Sector | Phase | E | T | M | U | L |
|---|---|---|---|---|---|---|---|
| O3 | Finance | Baseline | 0.669 | 0.652 | 0.663 | 0.592 | 0.535 |
| O3 | Finance | Post | 0.359 | 0.880 | 0.601 | 0.737 | 0.250 |
| O4 | Finance | Baseline | 0.427 | 0.683 | 0.594 | 0.640 | 0.367 |
| O4 | Finance | Post | 0.305 | 0.854 | 0.535 | 0.783 | 0.196 |
| O5 | Healthcare | Baseline | 0.545 | 0.369 | 0.647 | 0.365 | 0.724 |
| O5 | Healthcare | Post | 0.242 | 0.790 | 0.459 | 0.671 | 0.173 |
| O6 | Healthcare | Baseline | 0.503 | 0.434 | 0.633 | 0.388 | 0.654 |
| O6 | Healthcare | Post | 0.204 | 0.842 | 0.432 | 0.701 | 0.130 |
| O7 | Healthcare | Baseline | 0.448 | 0.477 | 0.616 | 0.417 | 0.581 |
| O7 | Healthcare | Post | 0.172 | 0.843 | 0.431 | 0.711 | 0.110 |
| O8 | Manufacturing | Baseline | 0.628 | 0.321 | 0.622 | 0.200 | 0.859 |
| O8 | Manufacturing | Post | 0.301 | 0.749 | 0.456 | 0.555 | 0.248 |
| O9 | Manufacturing | Baseline | 0.684 | 0.288 | 0.636 | 0.185 | 0.891 |
| O9 | Manufacturing | Post | 0.310 | 0.767 | 0.432 | 0.562 | 0.237 |
| O10 | Manufacturing | Baseline | 0.494 | 0.415 | 0.584 | 0.265 | 0.724 |
| O10 | Manufacturing | Post | 0.215 | 0.796 | 0.402 | 0.597 | 0.154 |
| O11 | Manufacturing | Baseline | 0.512 | 0.430 | 0.588 | 0.270 | 0.722 |
| O11 | Manufacturing | Post | 0.228 | 0.788 | 0.416 | 0.593 | 0.169 |
| O12 | Higher Education | Baseline | 0.605 | 0.368 | 0.567 | 0.319 | 0.745 |
| O12 | Higher Education | Post | 0.314 | 0.775 | 0.375 | 0.632 | 0.194 |
| O13 | Higher Education | Baseline | 0.503 | 0.368 | 0.546 | 0.345 | 0.684 |
| O13 | Higher Education | Post | 0.261 | 0.759 | 0.367 | 0.648 | 0.163 |
| O14 | Public Administration | Baseline | 0.668 | 0.274 | 0.646 | 0.201 | 0.887 |
| O14 | Public Administration | Post | 0.344 | 0.756 | 0.438 | 0.583 | 0.255 |
| O15 | Public Administration | Baseline | 0.612 | 0.320 | 0.631 | 0.204 | 0.855 |
| O15 | Public Administration | Post | 0.306 | 0.756 | 0.436 | 0.582 | 0.233 |

*Note. E (Exposure) and M (Motivation) are likelihood enhancers; T (Traceability) and U (Systems Update) are mitigating components.* $L = \sigma(\alpha \ln(\varepsilon+E) + \beta \ln(\varepsilon+M) - \gamma \ln(\varepsilon+T) - \delta \ln(\varepsilon+U))$ *under the reference parameterization* $\alpha = \beta = \gamma = \delta = 1$, $\varepsilon = 10^{-6}$. *All scores are rounded to three decimal places. Post-intervention values correspond to the first full monthly window after the documented intervention time* $\tau_0$ *for each organization.*

## 5.4. Feedback loop

The results demonstrate that the framework operates as an iterative refinement loop in which A5 outputs drive each successive intervention cycle and updated telemetry from that intervention is reintroduced into the cyber-exposure profile for the next window's scoring.

At each window, A5 produces a ranked control list by applying the four-step procedure described in the Methodology to the current component scores and graph state. In the first intervention cycle, the component leverage ranking in Step 1 consistently identified Traceability (T) and Exposure (E) as the primary targets across all 15 organizations, because these components showed the largest normalized deficits — T was consistently far below its policy threshold due to low logging coverage, and E was elevated by high rates of unregistered devices and privileged-user concentration. The Step 2 metric deficit analysis within those components pinpointed the specific metrics driving each condition: for Traceability, the dominant deficits were in log coverage (devices online / registered devices) and authenticated-user monitoring; for Exposure, the dominant deficits were in registered-device ratio and privileged-user concentration. The Step 3 graph query then retrieved controls mapped to those conditions through the MITIGATED_BY edge structure: for Traceability deficits, D3FEND techniques in the Log Management, Credential Activity Analysis, and User Behavior Analysis families; for Exposure deficits, techniques in the Asset Inventory, Account Management, and Network Traffic Filtering families. Step 4 scored these controls by coverage, severity, and cross-component impact, ranking actions that simultaneously reduced Exposure and improved Traceability — such as enrolling unregistered devices in an MDM with mandatory logging — above single-dimension controls.

This ranking was consistent with the interventions actually implemented across the study organizations, providing one form of face validity for the A5 mechanism: the cross-component controls (device registration with telemetry enrolment, privileged-account review with correlation-rule update) that Step 4 ranked highest were also the interventions associated with the largest observed improvements in both L and in the empirical outcomes reported in Tables 2 and 3.

In subsequent cycles, each intervention altered the graph state — new asset registrations extended the OWNS edge set, new log sources created LogSource nodes and DetectionRule edges, patched vulnerabilities removed EXPLOITS edges — and the next window's A5 query traversed the updated profile. This made it possible to confirm whether a prior recommendation had been acted upon (the addressed metric deficit closed) or whether the condition persisted (the same control re-appeared in the new cycle's ranked output). The most effective interventions were those that closed the deficit completely, removing the affected assets or conditions from the high-deficit set and allowing A5 to surface the next-priority control target. Organizations that implemented controls partially — for example, enrolling devices in the MDM but not configuring log forwarding — saw the Traceability deficit persist in the subsequent cycle's Step 2 ranking, and the A5 output re-recommended the missing configuration step with an updated coverage score reflecting the partially closed gap. This iterative diagnostic behaviour is the operational function of A5: it does not merely recommend controls once, but tracks whether measured conditions actually improve and re-prioritizes accordingly each window.

# 6. Discussion

## 6.1. Main findings and interpretation

The results indicate that the proposed framework can be instantiated consistently across organizations with different infrastructures, sectors, and defensive baselines, to obtain a cyber-exposure profile integrating Exposure, Traceability, Motivation, and Systems Update scoring. Those scores were mapped to a bounded likelihood indicator that could be interpreted longitudinally and compared across cases. The principal contribution is not the constructs themselves — which reflect established security engineering concepts — but the formal specification of the artifact chain that converts heterogeneous organizational evidence into a reproducible, bounded indicator using those constructs.

This operationalization is especially relevant in cybersecurity because incidents are underreported, technical environments change rapidly, and attacker behaviour is adaptive. Under those conditions, historical counts by themselves are insufficient for decision support. The framework addresses that limitation by measuring the organization's current cyber condition through four interpretable dimensions: Exposure, which captures attack surface and exploitable opportunity; Traceability, which captures observability and attribution capacity; Motivation, which captures attacker incentive and target attractiveness; and Systems Update, which captures patch posture, maintainability, and update currency. This evidence-based, continuously updatable approach is particularly relevant in environments where threat conditions evolve rapidly, including contexts where adversaries leverage automated or AI-assisted attack techniques.

## 6.2. Practical implications

The framework offers at least three practical advantages, each receiving preliminary empirical support. First, it provides a transparent method for converting heterogeneous evidence into a comparable likelihood-related output — a property demonstrated across 15 organizations but not yet tested at scale or against alternative methods.

One important caveat concerns the 35% mean reduction in incident frequency. Because the interventions recommended by the framework — registering unmanaged devices, reducing privileged-user concentration, expanding logging — directly alter the metric inputs that produce the likelihood score, score improvement is expected whenever the recommended controls are implemented. This makes it difficult, at the current sample size and without a comparison condition, to distinguish the specific discriminative value of the scoring formula from the general benefit of any structured security engagement. The pre-intervention directional consistency check ($r = .736$, $p = .002$) addresses only the first mechanism, showing that the pre-intervention score orders organizations consistently with observed incident frequency before any controls are changed. Demonstrating added value beyond structured engagement per se requires a matched comparison group — recommended as the primary design feature of future evaluations.

Second, the framework improves traceability between technical observations and management decisions, since each component and metric can be traced to specific sources, queries, and control conditions. Third, it supports control determination and prioritization, because changes in the likelihood score can be linked back to concrete interventions such as expanding logging, registering unmanaged assets, reducing privilege concentration, or accelerating critical updates.

## 6.3. Methodological implications

The study shows the value of combining design-science principles, data-science workflows, graph-based integration, and case-based evaluation. A central criticism of many cyber-risk models is that they present formulas or high-level constructs without sufficient explanation of how underlying data are gathered, transformed, and linked to outcomes. The framework addresses that weakness through explicit artifacts, documented computational traceability, and a case instantiation procedure that follows the computation chain from source evidence to graph integration, metric computation, likelihood scoring, and outcome evaluation.

A specific methodological contribution is the articulation of the cyber-exposure profile as a formally induced, time-bounded subgraph rather than as an informal narrative or a flat attribute table. This formalization matters because it makes the profile queryable, versioned, and reproducible: each metric in Table 1 can be traced to a specific graph query over a specific evidence bundle. It also makes iterative updating natural — adding new telemetry, a revised vulnerability scan, or a new intervention record produces a new induced subgraph over the same schema, enabling comparison across time windows without ad hoc re-engineering. The use of the log-additive/logistic formulation for likelihood scoring, while not unique to this study, is justified here by four explicit design requirements (monotonicity, boundedness, robustness, auditability) rather than by convention, and the robustness check against a linear alternative (Spearman $r_s = .97$) confirms that the functional form choice is not a material driver of the reported ordinal results at this cohort size.

## 6.4. Limitations

The study uses a purposive multiple-case design rather than a statistically representative sample, limiting statistical generalization. Incident counts and detection-and-response measures are influenced by reporting maturity, telemetry coverage, and organizational thresholds for escalation. Organizations that substantially expanded logging coverage during the intervention period — most notably manufacturing and public-administration cases, where baseline coverage was below 20% and post-intervention coverage exceeded 93% — were recording incidents through a materially different detection funnel before and after intervention; some portion of the observed incident reduction therefore reflects stabilization of that funnel to a new higher baseline rather than solely genuine security improvement. This parsing is consistent with the finance sector pattern: those organizations had the highest baseline logging coverage and the smallest incident reductions (4–9%), interpretable as both a narrower margin for improvement and a more stable detection funnel less affected by this confound. Future studies should record detected-but-not-escalated events separately so that the detection funnel effect can be modelled explicitly.

The two component-level correlations in Table 3 ($r = .751$ and $r = .700$) carry wide confidence intervals at $N = 15$, spanning approximately 52 and 60 percentage points respectively, and should be read as directional associations requiring replication in a larger cohort before conclusions about the magnitude of these relationships can be drawn. A further limitation is the absence of a comparison condition — organizations using an alternative framework or no structured framework — which prevents isolating the specific effect of the scoring methodology from the effect of any structured security engagement. A "measure, identify gaps, remediate" cycle would be expected to produce some improvement regardless of the particular measurement instrument used. Future evaluations should include matched comparison organizations or, where feasible, a stepped-wedge design in which different organizations enter the intervention at staggered points.

### 6.5. Future research

Future research can build on this work in several directions. One path is prospective validation in larger and more diverse cohorts, including organizations with different maturity profiles and different combinations of IT, cloud, and OT environments; a stepped-wedge design with matched comparison organizations would provide the strongest basis for attributing observed gains to the framework's scoring logic rather than to the general benefit of structured security engagement. A second path is methodological extension through automated data acquisition, continuous graph updates, and near-real-time likelihood re-computation, reducing the overhead of evidence-bundle assembly and enabling shorter observation windows. A third path is richer modelling, including machine-learning-assisted estimation of sensitivity coefficients from historical outcome data, alternative weighting schemes derived through stakeholder elicitation, or probabilistic graphical formulations that could complement the current bounded score with explicit uncertainty intervals rather than point estimates. Another important direction is deeper incorporation of sectoral and regulatory context into the Motivation dimension, particularly in domains such as healthcare, finance, public administration, and industrial control environments, where attacker incentives and operational dependencies differ substantially and where regulatory requirements provide additional observable signals for both Motivation and Systems Update scoring.

## 7. Conclusions

This article proposes a framework for operationalizing cybersecurity incident likelihood when historical data are incomplete, inconsistently formatted, or poorly comparable across organizations. The core idea is to build an operational cyber-exposure profile that integrates heterogeneous cyber evidence and converts it into four normalized components — Exposure, Traceability, Motivation, and Systems Update — for mapping to a bounded likelihood indicator.

The contribution is primarily methodological: a transparent chain links source acquisition, graph construction, organization-specific profiling, metric computation, normalization, likelihood scoring, and control-oriented interpretation. The chain moves likelihood assessment away from opaque expert-only scoring and toward an evidence-based procedure that can be recalculated as conditions change. The four constructs are not claimed as novel discoveries — they reflect established categories in security engineering practice — but their value lies in their precise operationalization within the pipeline and their empirical grounding in a large incident corpus.

The empirical evaluation across 15 organizations provides preliminary support for the usefulness of the approach. The reported reductions in incident frequency and detection/response time, together with the artifact-to-outcome traceability demonstrated in the case study, suggest that the framework captures meaningful aspects of organizational cyber condition. More broadly, the study provides preliminary evidence consistent with the proposition that cybersecurity likelihood can be treated as a measurable and updateable condition, that graph-based integration can improve the transparency of risk reasoning, and that a metric-driven model can connect risk assessment more directly to control prioritization and resource allocation.

The framework should be read as a structured basis for evidence-based likelihood assessment in cybersecurity: useful because it makes the score explainable, reproducible, and actionable, while remaining cautious about claims of universal predictive probability. Confirming its discriminative value, transferability across sectors, and calibration performance relative to alternatives remains the work of future evaluations conducted at larger scale and with appropriate comparison conditions.

**Author Contributions:** Conceptualization, Pablo Corona Fraga, Dr. Vanessa Diaz Rodriguez, and Dr. Jesús Manuel Niebla Zatarain; methodology, P.C.F.; software, P.C.F.; validation, P.C.F., V.D.R., and J.M.N.Z.; formal analysis, V.D.-R., Gabriel Sánchez-Pérez, J.M.N.Z., and Dr. Edward J. Humphreys; investigation, P.C.F.; resources, P.C.F.; data curation, P.C.F.; writing—original draft preparation, P.C.F. and V.D.R.; writing—review and editing, V.D.-R., G.S.P.,

J.M.N.Z., and E.J.H. All authors have read and agreed to the published version of the manuscript.

**Funding:** No funding was received.

**Institutional Review Board Statement:** Not applicable.

**Informed Consent Statement:** Not applicable.

**Data Availability Statement:** The original data presented in the study are openly available: VERIS Community Database (VCDB) https://github.com/vz-risk/VCDB; Hackmageddon cyber-attacks timeline collection https://www.hackmageddon.com/category/security/cyber-attacks-timeline/ (accessed on 17 April 2026).

**Conflicts of Interest:** The authors declare no conflict of interest.